\documentstyle[epsf,graphics,epsfig]{mn}

\def\mpcoh{\,h^{-1}\,{\rm Mpc}}

\def\spose#1{\hbox to 0pt{#1\hss}}

\def\simgt{\mathrel{\lower0.6ex\hbox{$\buildrel {\textstyle >}
 \over {\scriptstyle \sim}$}}}
\def\simlt{\mathrel{\lower0.6ex\hbox{$\buildrel {\textstyle <}
 \over {\scriptstyle \sim}$}}}

\def\m@th{\mathsurround=0pt }
\def\eqalign#1{\null\,\vcenter{\openup1\jot \m@th
 \ialign{\strut\hfil$\displaystyle{##}$&$\displaystyle{{}##}$\hfil
 \crcr#1\crcr}}\,}

\def\japsub{\scriptscriptstyle\rm}

\def\wm{{\omega_m}} 
\def\wb{{\omega_b}}
\def\Oml{{ \Omega_{\Lambda} }}
\def\aeq{{ a_{\rm eq} }}
\def\bfrac{{ \Omega_b/\Omega_m }}

\font\japit = cmti10 at 11truept

\title[Parameter constraints for flat cosmologies from CMB and 2dFGRS
power spectra]
{\vglue-3.0truecm
\centerline{\japit Submitted for publication in Monthly Notices of the R.A.S.}
\vglue 2.5truecm\noindent
Parameter constraints for flat cosmologies from CMB and 2dFGRS power
spectra}

\begin{document}

\author[W.J.~Percival et al.]{
\parbox[t]{\textwidth}{
Will J.\ Percival$^1$,
Will Sutherland$^1$,
John A.\ Peacock$^1$,
Carlton M.\ Baugh$^2$,
Joss Bland-Hawthorn$^3$,
Terry Bridges$^3$, 
Russell Cannon$^3$, 
Shaun Cole$^2$, 
Matthew Colless$^4$, 
Chris Collins$^5$, 
Warrick Couch$^6$, 
Gavin Dalton$^{7,8}$,
Roberto De Propris$^6$,
Simon P.\ Driver$^9$, 
George Efstathiou$^{10}$, 
Richard S.\ Ellis$^{11}$, 
Carlos S.\ Frenk$^2$, 
Karl Glazebrook$^{12}$, 
Carole Jackson$^4$,
Ofer Lahav$^{10}$, 
Ian Lewis$^3$, 
Stuart Lumsden$^{13}$, 
Steve Maddox$^{14}$,
Stephen Moody$^9$,
Peder Norberg$^2$,
Bruce A.\ Peterson$^4$, 
Keith Taylor$^3$ (The 2dFGRS Team)}
\vspace*{6pt} \\ 
$^1$Institute for Astronomy, University of Edinburgh, Royal Observatory, 
       Blackford Hill, Edinburgh EH9 3HJ, UK \\
$^2$Department of Physics, University of Durham, South Road, Durham DH1 3LE, UK \\
$^3$Anglo-Australian Observatory, P.O.\ Box 296, Epping, NSW 2121,
    Australia\\  
$^4$Research School of Astronomy \& Astrophysics, The Australian 
    National University, Weston Creek, ACT 2611, Australia \\
$^5$Astrophysics Research Institute, Liverpool John Moores University,  
    Twelve Quays House, Birkenhead, L14 1LD, UK \\
$^6$Department of Astrophysics, University of New South Wales, Sydney, 
    NSW 2052, Australia \\
$^7$Department of Physics, University of Oxford, Keble Road, Oxford OX1 3RH, UK \\
$^8$Space Science and Technology Division, Rutherford Appleton
    Laboratory, Chilton, Didcot, OX11 0QX, UK \\
$^9$School of Physics and Astronomy, University of St Andrews, North
    Haugh, St Andrews, Fife, KY6 9SS, UK \\
$^{10}$Institute of Astronomy, University of Cambridge, Madingley Road,
       Cambridge CB3 0HA, UK \\
$^{11}$Department of Astronomy, Caltech, Pasadena, CA 91125, USA \\
$^{12}$Department of Physics \& Astronomy, Johns Hopkins University,
       Baltimore, MD 21218-2686, USA \\
$^{13}$Department of Physics, University of Leeds, Woodhouse Lane,
       Leeds, LS2 9JT, UK \\
$^{14}$School of Physics \& Astronomy, University of Nottingham,
       Nottingham NG7 2RD, UK \\
}

\date{Accepted. Received ; in original form}

\maketitle

\begin{abstract}
We constrain flat cosmological models with a joint likelihood analysis
of a new compilation of data from the cosmic microwave background
(CMB) and from the 2dF Galaxy Redshift Survey (2dFGRS). Fitting the
CMB alone yields a known degeneracy between the Hubble constant $h$
and the matter density $\Omega_m$, which arises mainly from preserving
the location of the peaks in the angular power spectrum. This
`horizon-angle degeneracy' is considered in some detail and shown to
follow a simple relation $\Omega_m h^{3.4} = {\rm constant}$. Adding
the 2dFGRS power spectrum constrains $\Omega_m h$ and breaks the
degeneracy.  If tensor anisotropies are assumed to be negligible, we
obtain values for the Hubble constant $h=0.665\pm0.047$, the matter
density $\Omega_m=0.313\pm0.055$, and the physical CDM and baryon
densities $\Omega_c h^2 = 0.115 \pm 0.009, \Omega_b h^2 = 0.022\pm
0.002$ (standard rms errors). Including a possible tensor component
causes very little change to these figures; we set an upper limit to
the tensor-to-scalar ratio of $r<0.7$ at 95\% confidence.  We then
show how these data can be used to constrain the equation of state of
the vacuum, and find $w<-0.52$ at 95\% confidence. The preferred
cosmological model is thus very well specified, and we discuss the
precision with which future CMB data can be predicted, given the model
assumptions. The 2dFGRS power-spectrum data and covariance matrix, and
the CMB data compilation used here, are available from {\tt
http://www.roe.ac.uk/{\tt\char'176}wjp/}.
\hfill\break \hfill\break {\bf keywords}: large-scale structure of
Universe, cosmic microwave background, cosmological parameters
\end{abstract}

\strut\vfill\eject
\strut\vfill\eject

\section{introduction}

The 2dF Galaxy Redshift Survey (2dFGRS; see e.g. Colless et~al. 2001)
has mapped the local Universe in detail.  If the galaxy distribution
has Gaussian statistics and the bias factor is independent of scale,
then the galaxy power spectrum should contain
all of the available information about the seed perturbations of
cosmological structure: it is statistically complete in the linear
regime. The power spectrum of the data as of early 2001 was presented
in Percival et~al. (2001), and was shown to be consistent with recent
cosmic microwave background (CMB) and nucleosynthesis results.

In Efstathiou et~al. (2002) we combined the 2dFGRS power spectrum with
recent CMB datasets in order to constrain the cosmological model (see
also subsequent work by Lewis \& Bridle 2002).  Considering a wide
range of possible assumptions, we were able to show that the universe
must be nearly flat, requiring a non-zero cosmological constant
$\Lambda$.  The flatness constraint was quite precise ($|1-\Omega_{\rm
tot}| < 0.05$ at 95\% confidence); since inflation models usually
predict near-exact flatness ($\vert 1 - \Omega_{\rm tot} \vert <
0.001$; e.g. Section 8.3 of Kolb \& Turner 1990), there is strong
empirical and theoretical motivation for considering only the class of
exactly flat cosmological models.  The question of which flat
universes match the data is thus an important one to be able to
answer.  Removing spatial curvature as a degree of freedom also has
the practical advantage that the space of cosmological models can be
explored in much greater detail.  Therefore, throughout this work we
assume a universe with baryons, CDM and vacuum energy summing to
$\Omega_{\rm tot} = 1$ (cf.  Peebles 1984; Efstathiou, Sutherland \&
Maddox 1990).

In this work we also assume that the initial fluctuations were
adiabatic, Gaussian and well described by power law spectra. We
consider models with and without a tensor component, which is allowed
to have slope and amplitude independent of the scalar component.
Recent Sudbury Neutrino Observatory (SNO) measurements (Ahmad et~al.
2002) are most naturally interpreted in terms of three neutrinos of
cosmologically negligible mass ($\simlt 0.05$~eV, as opposed to
current cosmological limits of order 2~eV -- see Elgaroy et~al.
2002). We therefore assume zero neutrino mass in this analysis.  In
most cases, we assume the vacuum energy to be a `pure' cosmological
constant with equation of state $w \equiv p / \rho c^2 = -1$, except
in Section~\ref{sec:quin} where we explore $w > -1$.

In Section~\ref{sec:CMBdata} we use a compilation of recent CMB
observations (including data from VSA (Scott et~al. 2002) and CBI
(Pearson et~al. 2002) experiments) to determine the maximum-likelihood
amplitude of the CMB angular power spectrum on a convenient grid,
taking into account calibration and beam uncertainties where
appropriate. This compression of the data is designed to speed the
analysis presented here, but it should be of interest to the community
in general.

In Section~\ref{sec:models} we fit to both the CMB data alone, and CMB
+ 2dFGRS.  Fits to CMB data alone reveal two well-known primary
degeneracies. For models including a possible tensor component, there
is the tensor degeneracy (Efstathiou 2002) between increasing tensors,
blue tilt, increased baryon density and lower CDM density.  For both
scalar-only and with-tensor models, there is a degeneracy related to
the geometrical degeneracy present when non-flat models are
considered, arising from models with similar observed CMB peak
locations (cf. Efstathiou \& Bond 1999).  In Section~\ref{sec:horizon}
we discuss this degeneracy further and explain how it may be easily
understood via the horizon angle, and described by the simple relation
$\Omega_m h^{3.4} = {\rm constant}$.

Section~\ref{sec:quin} considers a possible extension of our standard
cosmological model allowing the equation of state parameter $w$ of the
vacuum energy component to vary. By combining the CMB data, the 2dFGRS
data, and an external constraint on the Hubble constant $h$, we are
able to constrain $w$. Finally, in Section~\ref{sec:cls}, we discuss
the range of CMB angular power spectral values allowed by the present
CMB and 2dFGRS data within the standard class of flat models.

\section{The CMB data}  \label{sec:CMBdata}

\begin{table} 

  \caption{ Best-fit relative power calibration corrections for the
  experiments considered are compared to expected rms errors. In
  addition, we recover a best fit beam error for BOOMERaNG of +0.4\%,
  measured relative to the first data point in the set, and +0.07\%
  for Maxima. \label{tab:calib} }

  \centering \begin{tabular}{lcc} \hline
	& \multicolumn{2}{c}{power calibration error} \\
  experiment & best-fit (\%) & rms (\%) \\
  \hline

 BOOMERaNG & $-$13.5 & 20 \\
 Maxima    & +1.6  & 8  \\
 DASI      & +0.9  & 8  \\
 VSA       & $-$0.3  & 7  \\
 CBI       & +0.7  & 10 \\
  \hline
  \end{tabular}

\end{table}

\begin{figure*}
  \setlength{\epsfxsize}{0.65\textwidth} 
  \centerline{\epsfbox{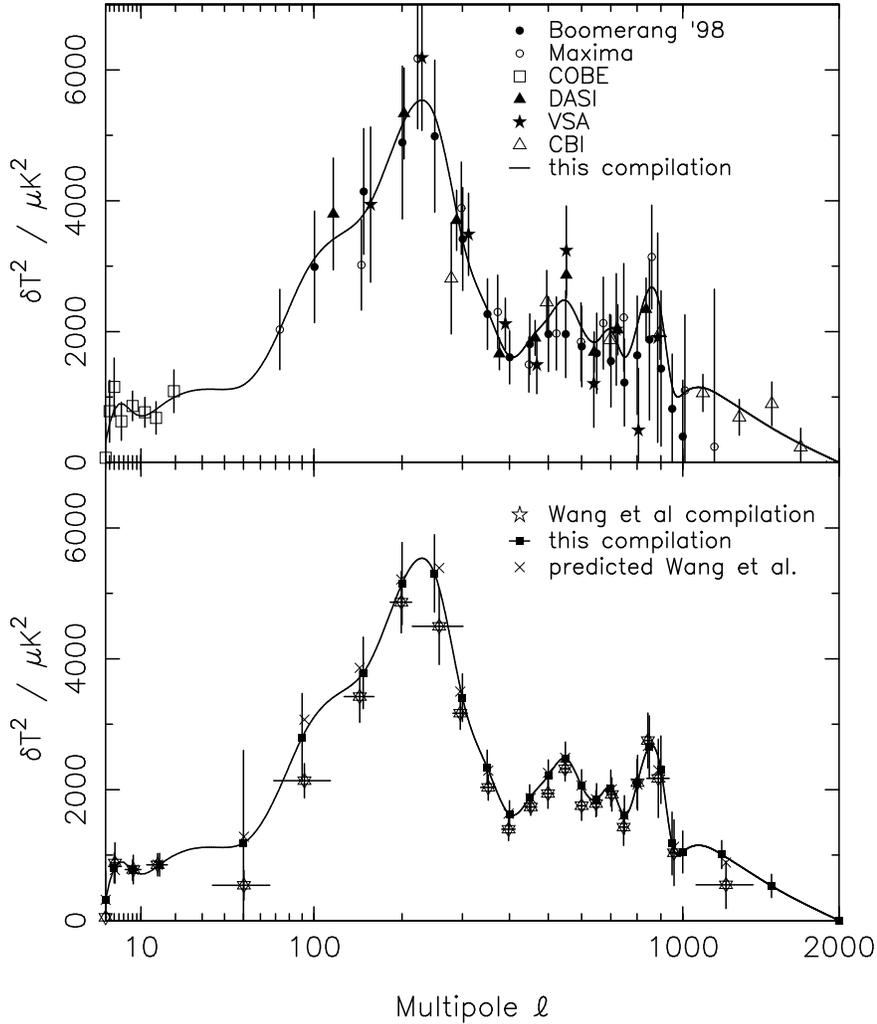}} 

  \caption{Top panel: the compilation of recent CMB data used in our
  analysis (see text for details). The solid line shows the result of
  a maximum-likelihood fit to these data allowing for calibration and
  beam uncertainty errors in addition to intrinsic errors. Each
  observed data set has been shifted by the appropriate best-fit
  calibration and beam correction. Bottom panel: the solid line again
  shows our maximum-likelihood fit to the CMB power spectrum now
  showing the nodes (the points at which the amplitude of the power
  spectrum was estimated) with approximate errors calculated from the
  diagonal elements of the covariance matrix (solid squares). These
  data are compared with the compilation of Wang et~al. (2002) (stars)
  and the result of convolving our best fit power with the window
  function of Wang et~al. (crosses). In order to show the important
  features in the CMB angular power spectrum plots we present in this
  paper we have chosen to scale the x-axis by $(\log\ell)^{2.5}$.
  \label{fig:CMB_data}}

\end{figure*}

Recent key additions to the field of CMB observations come from the
VSA (Scott et~al. 2002), which boasts a smaller calibration error than
previous experiments, and the CBI (Pearson et~al. 2002, Mason et~al.
2002), which has extended observations to smaller angles (larger
$\ell$'s). These data sets add to results from BOOMERaNG (Netterfield
et~al. 2002), Maxima (Lee et~al. 2001) and DASI (Halverson et~al.
2002), amongst others. Rather than compare models to each of these
data sets individually, it is expedient to combine the data prior to
analysis. This combination often has the advantage of allowing a
consistency check between the individual data sets (e.g. Wang et~al.
2002). However, care must be taken to ensure that additional biases
are not introduced into the compressed data set, and that no important
information is lost. 

In the following we consider COBE, BOOMERaNG, Maxima, DASI, VSA and
CBI data sets. The\\ BOOMERaNG data of Netterfield et~al. (2002) and the
Maxima data of Lee et~al. (2002) were used assuming the data points
were independent, and have window functions well described by
top-hats. The $\ell<2000$ CBI mosaic field data were used assuming
that the only significant correlations arise between neighbouring
points which are anti-correlated at the $16\%$ level as discussed in
Pearson et~al. (2002). Window functions for these data were assumed to
be Gaussian with small negative side lobes extending into neighbouring
bins approximately matched to figure~11 of Pearson et~al. (2002). We
also consider the VSA data of Scott et~al. (2002), the DASI data of
Halverson et~al. (2002), and the COBE data compilation of Tegmark
et~al. (1996), for which the window functions and covariance matrices
are known, where appropriate. The calibration uncertainties used are
presented in Table~\ref{tab:calib}, and the data sets are shown in
Fig.~\ref{fig:CMB_data}. In total, there are 6 datasets, containing 68
power measurements.

In order to combine these datasets, we have fitted a model for the
true underlying CMB power spectrum, consisting of power values at a
number of $\ell$ values (or nodes).  Between these nodes we
interpolate the model power spectrum using a smooth Spline3 algorithm
(Press et~al. 1992). The assumption of smoothness is justified because
we aim to compare CMB data with CDM models calculated using CMBFAST
(Seljak \& Zaldarriaga 1996).  Internally, this code evaluates the CMB
power spectrum only at a particular set of $\ell$ values, which are
subsequently Spline3 interpolated to cover all multipoles.  It is
therefore convenient to use as our parameters the CMB power values at
the same nodes used by CMBFAST in the key regime
$150\le\ell\le1000$. By using the same smoothing algorithm and nodes
for our estimate of the true power spectrum, we ensure that no
additional assumptions are made in the data compilation compared with
the models to be tested. For $\ell<150$ and $\ell>1000$ the data
points are rather sparsely distributed, and we only selected a few
$\ell$ values at which to estimate the power.  The best-fit amplitude
of the power spectrum at an extra node at $\ell=2000$ was determined
in our fit to the observed CMB data, in order that the shape of the
interpolated curve around $\ell=1500$ had the correct form. This was
subsequently removed from the analysis, and models and data were only
compared for $\ell\le1500$.  In addition to requiring no interpolation
in CMBFAST, this method of compression has a key advantage for our
analysis. Normally, CMB data are expressed as bandpowers, in which one
specifies the result of convolving the CMB power spectrum with some
kernel. This remains true of some previous CMB data compilations
(e.g. Wang et al. 2002).  In contrast, we estimate the true power
spectrum at a given $\ell$ directly, so that no convolution step is
required. This means that parameter space can be explored more quickly
and efficiently.

Given a set of nodal values, we form an interpolated model power
spectrum, convolve with the window function of each observed data
point and maximized the Likelihood with respect to the nodal values
(assuming Gaussianity -- see Bond, Jaffe \& Knox 2000 for a
discussion of the possible effect of this approximation).  Calibration
errors and beam uncertainties were treated as additional independent
Gaussian parameters, and were combined into the final likelihood, as
well as being used to correct the data. The resulting best fit
calibration and beam errors are compared to the expected rms values in
Table~\ref{tab:calib}.

\begin{table} 

  \caption{ Recovered best-fit power spectrum values with rms values
  given the 6 data sets analysed. \label{tab:data} }

  \centering \begin{tabular}{ccc} \hline
  $\ell$ & $\delta T^2$ / $\mu K^2$ & rms error / $\mu K^2$ \\
  \hline
	2   &     314  & 443  \\
	4   &     803  & 226  \\
	8   &     770  & 156  \\
	15  &     852  & 174  \\
	50  &     1186 & 1414 \\
	90  &     2796 & 673  \\
	150 &     3784 & 546  \\
	200 &     5150 & 627  \\
	250 &     5306 & 590  \\
	300 &     3407 & 364  \\
	350 &     2339 & 265  \\
	400 &     1627 & 205  \\
	450 &     1873 & 202  \\
	500 &     2214 & 240  \\
	550 &     2479 & 249  \\
	600 &     2061 & 245  \\
	650 &     1849 & 244  \\
	700 &     2023 & 274  \\
	750 &     1614 & 295  \\
	800 &     2089 & 373  \\
	850 &     2654 & 475  \\
	900 &     2305 & 515  \\
	950 &     1178 & 480  \\
	1000 &    1048 & 320  \\
	1200 &    1008 & 214  \\
	1500 &    530  & 178  \\
  \hline
  \end{tabular}

\end{table}

In agreement with Wang et~al. (2002), we find a negative best fit
BOOMERaNG calibration correction (13\% in power), caused by matching
data sets in the regime $300<\ell<500$. Applying this correction
(included in the
data points in Fig.~\ref{fig:CMB_data}) 
slightly decreases the amplitude of the first
peak. Nevertheless, our combined power values are systematically higher
than in the compilation of Wang et al. (see the lower panel of Fig.~1).
This derives partly from the inclusion of extra data, but also results
from a bias in the analysis method of Wang et al. 
They use the observed power values to estimate the error in the
data, rather than the true power at that multipole (which
we estimate from our model). A low observational point is thus
given a spuriously low error, and this is capable of biasing the
averaged data to low values.

The final best fit power spectrum amplitudes given the 6 data sets
analysed are presented in Table~\ref{tab:data}, with the corresponding
$\ell$-values of the nodes and rms errors. Formally this fit gave
$\chi^2_{\rm min}=31.9$, given 34 degrees of freedom (there are 68
data points, and we estimate 27 power spectrum values, 5 calibration
and 2 beam corrections). This result demonstrates that the different
data sets are broadly consistent after allowing for calibration and
beam uncertainty. The Hessian matrix of the likelihood provides an
estimate of the inverse covariance matrix for the power spectrum
estimates. This was calculated numerically and is available, together
with the averaged data, from 
{\tt http://www.roe.ac.uk/{\tt\char'176}wjp/}. As emphasised previously,
these are estimates of the true power at the $\ell$ values given and
therefore do not require window functions. In the following Section we
use these CMB results to constrain flat cosmological models.

\begin{figure*}
  \setlength{\epsfxsize}{0.95\textwidth} \centerline{\epsfbox{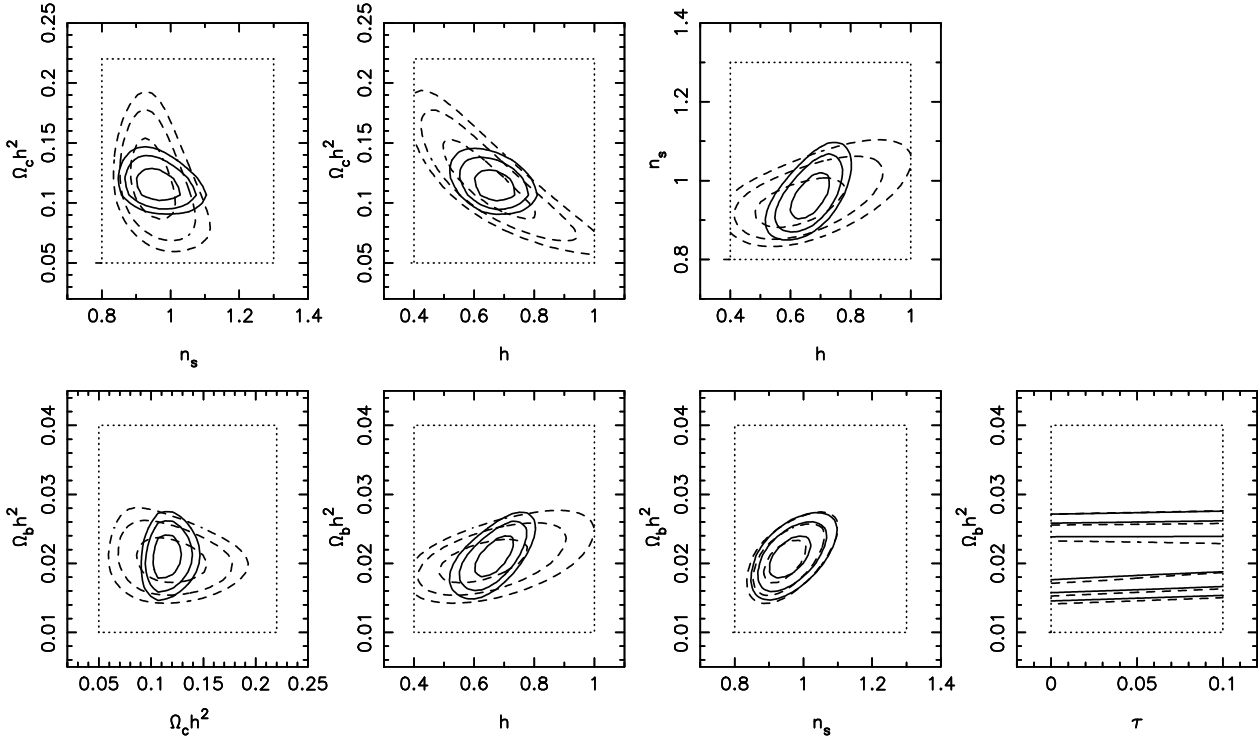}}

  \caption{Two parameter likelihood surfaces for scalar only
  models. Contours correspond to changes in the likelihood from the
  maximum of $2\Delta\ln{\cal L}=2.3, 6.0, 9.2$. Dashed contours are
  calculated by only fitting to the CMB data, solid contours by
  jointly fitting the CMB and 2dFGRS data. Dotted lines show the
  extent of the grid used to calculate the likelihoods.}

  \label{fig:like_sca}
\end{figure*}

\begin{figure*}
  \setlength{\epsfxsize}{0.95\textwidth} \centerline{\epsfbox{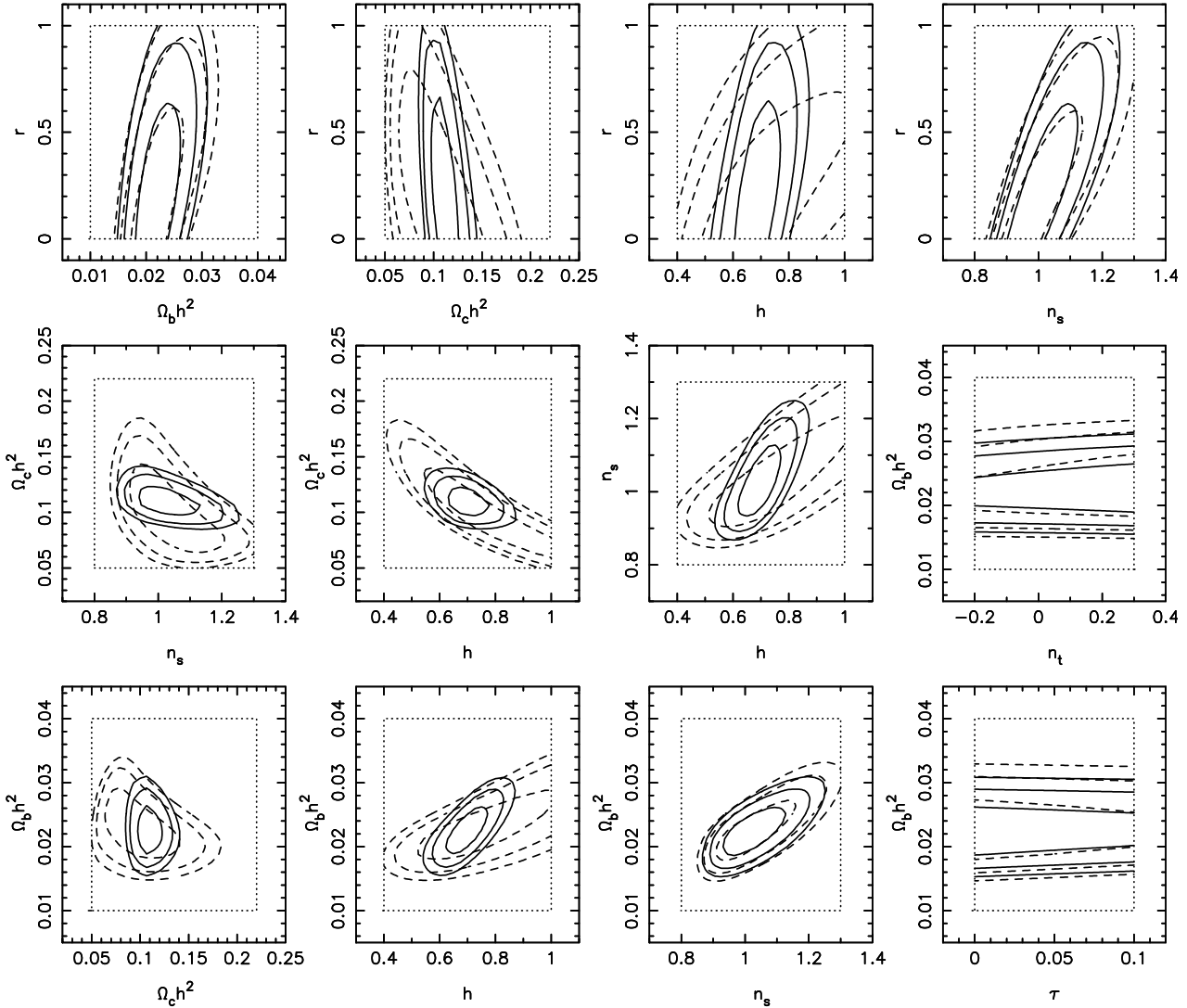}}

  \caption{As Fig.~\ref{fig:like_sca}, but now considering a wider
  class of models that possibly include a tensor component.}

  \label{fig:like_ten}
\end{figure*}

\section{cosmological models}  \label{sec:models}

\subsection{Parameter space}

In the following we parametrise flat cosmological models with seven
parameters (plus two amplitudes): these are the physical baryon
density\footnote{As usual, $\Omega_b, \Omega_c$ are the densities of baryons
\& CDM in units of the critical density, and $h$ is the Hubble
constant in units of $100 \,{\rm km\,s^{-1}\,Mpc^{-1}}$. `Derived'
parameters include the matter density $\Omega_m = \Omega_c +
\Omega_b$, and $\Oml = 1 - \Omega_m$. } $\Omega_bh^2$, the physical
CDM density $\Omega_ch^2$, the Hubble constant $h$, the optical depth
to the last scattering surface $\tau$, the scalar spectral index
$n_s$, the tensor spectral index $n_t$ and the tensor-to-scalar ratio
$r$. The tensor-to-scalar ratio $r$ is defined as in Efstathiou et~al.
(2002): the scalar and tensor $C_\ell$'s are normalized so that
\begin{eqnarray}
{1 \over 4 \pi}  \sum_{\ell=2}^{1000} (2\ell+1) \hat C^S_\ell & =&  (4\times10^{
-5})^2, \\
{1 \over 4 \pi}   \sum_{\ell=2}^{50}  (2\ell+1) \hat C^T_\ell & =&  (2\times 10^
{-5})^2. 
\end{eqnarray}
$C_\ell$ is then given by $C_\ell = Q^2( \hat C_\ell^S + r \hat
C_\ell^T)$, where $Q^2$ is the normalization constant. We marginalize
over both this and the amplitude of the 2dFGRS power spectra in order
to avoid complications caused by galaxy biasing and redshift space
distortions (Lahav et~al. 2002). 

\begin{table} 

  \caption{The distribution of parameters (defined in the text) in the
  $\sim2\times10^8$ flat cosmological models considered in this
  paper. The grid used was linear in each parameter between the limits
  given in order to simplify the marginalization assuming a uniform
  prior on each.  \label{tab:grid} }

  \centering \begin{tabular}{cccc} \hline
  parameter & min & max & grid size \\
  \hline

  $\Omega_bh^2$ &  0.01 & 0.04 & 25 \\
  $\Omega_ch^2$ &  0.05 & 0.22 & 25 \\
  $h$           &  0.40 & 1.00 & 25 \\
  $\tau$        &  0.00 & 0.10 & 2  \\
  $n_s$         &  0.80 & 1.30 & 25 \\
  $n_t$         & \llap{$-$}0.20 & 0.30 & 10 \\
  $r$           &  0.00 & 1.00 & 25 \\
  \hline
  \end{tabular}

\end{table}

CMB angular power spectra have been calculated using CMBFAST (Seljak
\& Zaldarriaga 1996) for a grid of $\sim2\times10^8$ models. For ease
of use, a uniform grid was adopted with a varying resolution in each
of the parameters (details of this grid are presented in
Table~\ref{tab:grid}). Likelihoods were calculated by fitting these
models to the reduced CMB data set presented in
Section~\ref{sec:CMBdata}. Similarly, large scale structure (LSS)
power spectra were calculated for the relevant models using the
fitting formula of Eisenstein \& Hu (1998), and were convolved with
the window function of the 2dFGRS sample, before being compared to the
2dFGRS data as in Percival et~al. (2001).

In order to constrain parameters, we wish to determine the probability
of each model in our grid given the available CMB and 2dFGRS
data. However, we can only easily calculate the probability of the
data given each model. In order to convert between these probabilities
using Bayes theorem, we need to adopt a prior probability for each
model or parameter. In this work, we adopt a uniform prior for the
parameters discussed above between the limits in Table~\ref{tab:grid}.
i.e. we assume that the prior probability of each model in the grid is
the same. Assuming a uniform prior for physically motivated parameters
is common in the field, although not often explicitly mentioned. Note
that the constraints placed by the current data are tight compared
with the prior, and that the biases induced by this choice are
therefore relatively small.

The likelihood distribution for a single parameter, or for two
parameters can be calculated by marginalizing the estimated
probability of the model given the data over all other
parameters. Because of the grid adopted in this work, we can do this
marginalization by simply averaging the ${\cal L}$ values calculated
at each point in the grid.

\begin{table*}  

  \caption{The recovered mean and root mean square (rms) error for
  each parameter, calculated by marginalizing over the remaining
  parameters. Results are presented for scalar-only and scalar+tensor
  models, and for CMB data only or CMB \& 2dFGRS power spectrum
  data. To reduce round-off error, means and rms errors are quoted to
  an accuracy such that the rms error has 2 significant figures. We
  also present constraints on some of the possible derived parameter
  combinations.  (Note that due to the marginalization, the
  maximum-likelihood values of `derived' parameters e.g. $\Omega_m$
  are not simply ratios of the ML values for each `independent'
  parameter) \label{tab:results} }

  \centering \begin{tabular}{cccccc} \hline
   & parameter & \multicolumn{2}{c}{results: scalar only} &
   \multicolumn{2}{c}{results: with tensor component} \\
              & & CMB & CMB+2dFGRS & CMB & CMB+2dFGRS \\
  \hline

  & $\Omega_bh^2$ & $0.0205\pm0.0022$ & $0.0210\pm0.0021$ & $0.0229\pm0.0031$ & $0.0226\pm0.0025$ \\
  & $\Omega_ch^2$ & $0.118\pm0.022$   & $0.1151\pm0.0091$ & $0.100\pm0.023$   & $0.1096\pm0.0092$ \\
  & $h$           & $0.64\pm0.10$     & $0.665\pm0.047$   & $0.75\pm0.13$     & $0.700\pm0.053$   \\
  \raisebox{1.5ex}[0pt]{using the}         
  & $n_s$         & $0.950\pm0.044$   & $0.963\pm0.042$   & $1.040\pm0.084$   & $1.033\pm0.066$   \\
  \raisebox{1.5ex}[0pt]{CMB data}         
  & $n_t$         & $-$               & $-$               & $0.09\pm0.16$     & $0.09\pm0.16$     \\
  \raisebox{1.5ex}[0pt]{compilation}
  & $r$           & $-$               & $-$               & $0.32\pm0.23$     & $0.32\pm0.22$     \\
  \raisebox{1.5ex}[0pt]{of Section~\ref{sec:CMBdata}}
  & $\Omega_m$    & $0.38\pm0.18$     & $0.313\pm0.055$   & $0.25\pm0.15$     & $0.275\pm0.050$   \\
  & $\Omega_mh$   & $0.226\pm0.069$   & $0.206\pm0.023$   & $0.174\pm0.063$   & $0.190\pm0.022$   \\
  & $\Omega_mh^2$ & $0.139\pm0.022$   & $0.1361\pm0.0096$ & $0.123\pm0.022$   & $0.1322\pm0.0093$ \\ 
  & $\bfrac$      & $0.152\pm0.031$   & $0.155\pm0.016$   & $0.193\pm0.048$   & $0.172\pm0.021$   \\

  \hline

  & $\Omega_bh^2$ & $0.0209\pm0.0022$ & $0.0216\pm0.0021$ & $0.0233\pm0.0032$ & $0.0233\pm0.0025$ \\
  & $\Omega_ch^2$ & $0.124\pm0.024$   & $0.1140\pm0.0088$ & $0.107\pm0.025$   & $0.1091\pm0.0089$ \\
  & $h$           & $0.64\pm0.11$     & $0.682\pm0.046$   & $0.74\pm0.14$     & $0.719\pm0.054$   \\
  \raisebox{1.5ex}[0pt]{using the}         
  & $n_s$         & $0.987\pm0.047$   & $1.004\pm0.047$   & $1.073\pm0.087$   & $1.079\pm0.073$   \\
  \raisebox{1.5ex}[0pt]{Wang et~al. (2002)}  
  & $n_t$         & $-$               & $-$               & $0.03\pm0.15$    & $0.03\pm0.15$    \\
  \raisebox{1.5ex}[0pt]{compilation} 
  & $r$           & $-$               & $-$               & $0.25\pm0.21$     & $0.27\pm0.20$     \\
  & $\Omega_m$    & $0.41\pm0.20$     & $0.296\pm0.051$   & $0.28\pm0.17$     & $0.261\pm0.048$   \\
  & $\Omega_mh$   & $0.240\pm0.076$   & $0.200\pm0.021$   & $0.189\pm0.071$   & $0.185\pm0.021$   \\
  & $\Omega_mh^2$ & $0.145\pm0.024$   & $0.1356\pm0.0092$ & $0.131\pm0.024$   & $0.1324\pm0.0088$ \\
  & $\bfrac$      & $0.149\pm0.033$   & $0.160\pm0.016$   & $0.186\pm0.049$   & $0.177\pm0.021$   \\

  \hline
  \end{tabular}

\end{table*}

In Fig.~\ref{fig:like_sca} we present two-parameter likelihood contour
plots (marginalized over the remaining parameters) for the subset of
scalar-only models i.e. $r$ fixed at $0$. For these scalar-only
models, we choose to plot $\tau$ only against $\Omega_bh^2$ as $\tau$
is poorly constrained by the CMB data, and has no degeneracies with
the other parameters.  In Fig.~\ref{fig:like_ten} we present
two-parameter likelihood contour plots (marginalized over the
remaining parameters) for models allowing a tensor component. The
spectral index of the tensor contribution is poorly constrained by the
CMB data so, as for $\tau$, we only show one plot with this parameter.

Figs~\ref{fig:like_sca}~\&~\ref{fig:like_ten} reveal two key
directions in parameter space that the CMB data have difficulty
constraining. When a tensor component is included, we have the tensor
degeneracy -- a trade-off between increasing tensors, increasing
$n_s$, increasing $\Omega_b h^2$ and decreasing $\Omega_c h^2$ (for
more detail see Efstathiou 2002). In addition, in both the scalar-only
and with-tensor cases, there is a degeneracy between $\Omega_c h^2$
and $h$, that results in the Hubble parameter $h$ being poorly
constrained by the CMB data alone. This degeneracy is discussed in
detail in the next Section.

We note that nearly all of the likelihood is contained well within our
prior regions, except for the case of tensor models with CMB-only data
in Fig~\ref{fig:like_ten}: here there is a region allowed by CMB
outside our priors with high tensor fraction, $h > 1, n_s \simeq 1.3,
\Omega_c h^2 \simeq 0.06$.  These parameters are ruled out by many
observations apart from 2dFGRS, so the
truncation is not a concern.

\subsection{Results}

The recovered mean and standard rms error calculated for each
parameter (except $\tau$ which is effectively unconstrained) are given
in Table~\ref{tab:results}.  What is striking is how well specified
many of the parameters are. 

The general features are as follows: changing from the Wang et~al. 
compilation to our compilation slightly shrinks the error bars (due to
VSA and CBI), but the central values are similar except for a slight
shift in $n_s$. Allowing tensors widens the error bars and causes
modest shifts in central values (the best fit has a zero tensor
fraction, but the fact that $r$ must be non-negative explains the
shifts).  The CMB data alone constrains $\Omega_b h^2$ and $n_s$ well
and $\Omega_c h^2$ quite well, but $\Omega_m$ and $h$ less well.
Adding the 2dFGRS data shrinks the errors on $\Omega_c h^2$, $h$ and
thus $\Omega_m$ and $\Omega_b/\Omega_m$ by more than a factor of 2.

The most restrictive case is the set of scalar-only models. These
yield $h=0.665$ with only a 7\% error, which is substantially better
than any other method. The matter density parameter comes out at
$\Omega_m = 0.313$, with a rather larger error of 18\%; errors on $h$
and $\Omega_m$ are anticorrelated so the physical matter density is
well determined, $\Omega_m h^2 = 0.136 \pm 7\%$.  We show below in
Section~\ref{sec:horizon} that this is because the CMB data measure
very accurately the combination $\Omega_m h^{3}$, so that an
accurate measurement of $\Omega_m$ requires $h$ to be known almost
exactly.

Moving from matter content to the fluctuation spectrum, the
scalar-only results give a tantalizing hint of red tilt, with
$n_s=0.963 \pm 0.042$.  Current data are thus within a factor of 2 of
the precision necessary to detect plausible degrees of tilt
(e.g. $n_s=0.95$ for $\lambda\phi^4$ inflation; 
see Section 8.3 of Liddle \& Lyth 2000).  Inflation of course
cautions against ignoring tensors, but it would be a great step
forward to rule out an $n_s=1$ scalar-only model. 

Including the possibility of tensors changes these conclusions only
moderately.  The errors on $h$ and $\Omega_m$ hardly alter, whereas
the error on $n_s$ rises to 0.066. The preferred model has $r=0$,
although this is rather poorly constrained. Marginalizing over the
other parameters, we obtain a 95\% confidence upper limit of $r<0.7$.
One way of ruling out the upper end of this range may be
to note that such tensor-dominated models predict a rather
low normalization for the present-day mass fluctuations,
as we now discuss.

\subsection{Normalization}

An advantage of the new CMB data included here is that the most
recent experiments have a rather small calibration uncertainty.
It is therefore possible to obtain precise values for the
overall normalization of the power spectrum. As usual,
we take this to be specified by $\sigma_8$, the rms density contrast
averaged over spheres of $8\mpcoh$ radius. For the scalar-only
grid of models shown in Fig. 2, this yields
\begin{equation}
\sigma_8 = (0.72 \pm 0.03 \pm 0.02)\; \exp \tau.
\end{equation}
The first error figure is the `theory error': the uncertainty
in $\sigma_8$ that arises because the conversion between the
observed $C_\ell$ and the present $P(k)$ depends on the
uncertain values of $\Omega_m$ etc. The second error figure
represents the uncertainty in the normalization
of the $C_\ell$ data (see Fig. 7). The total error in
$\sigma_8$ is the sum in quadrature of these two figures.

This result confirms with greater precision our previous conclusions
that allowed scalar-only models prefer a relatively low normalization
(Efstathiou et al. 2002; Lahav et al. 2002).  As discussed by Lahav et
al. (2002), a figure of $\sigma_8 = 0.72$ is consistent with the
relatively wide range of estimates from the abundance of rich
clusters, but is lower than the $\sigma_8 \simeq 0.9$ for
$\Omega_m\simeq 0.3$ preferred by weak lensing studies. The obvious
way to reconcile these figures is via the degenerate dependence of
$\sigma_8$ on $\tau$. The lowest plausible value for this is
$\tau=0.05$, corresponding to reionization at $z_r=8$ for the
parameters given here. To achieve $\sigma_8 = 0.9$ requires
$\tau=0.22$, or reionization at $z_r=22$, which is somewhat higher
than conventional estimates ($z_r < 15$; see e.g.  Loeb \& Barkana
2001). Additional evidence in this direction comes from the possible
first detection of Sunyaev-Zeldovich anisotropies at $\ell > 200$ by
the CBI (Mason et al. 2002). This signal is claimed to require
$\sigma_8\simeq 1$ (Bond et al. 2002), which would raise $z_r$ to
almost 30.  Further scrutiny of these independent estimates for
$\sigma_8$ will be required before one can claim evidence for first
object formation at extreme redshifts, but this is an exciting
possibility.

Finally, we note that these problems are sharpened if the
CMB power spectrum has a substantial tensor component.
As shown by Efstathiou et al. (2002), the model with the
maximal allowed tensor fraction ($r=0.6$) has a
normalization lower by a factor 1.18 than the best scalar-only
model. This pushes $z_r$ to almost 40 for $\sigma_8=1$, which starts to
become implausibly high, even given the large uncertainties
associated with the modelling of reionization.

\section{The horizon angle degeneracy}  \label{sec:horizon}

In this section we explore the degeneracy observed in
Figs.~\ref{fig:like_sca}~\&~\ref{fig:like_ten} between $\Omega_c h^2$
and $h$. This is related (but not identical) to the geometrical
degeneracy that exists when non-flat models are considered, and we now
show that it is very closely related to the location of the acoustic
peaks. Below, we first review the basics of the geometrical
degeneracy, secondly note why this is only weakly broken by the
flatness assumption, and thirdly give a simple heuristic argument why
this degeneracy approximately follows a contour of nearly constant $\Omega_m
h^{3}$.

\subsection{The geometrical degeneracy} 

The `geometrical degeneracy' in the CMB is well known (Zaldarriaga
et~al. 1997; Bond et~al. 1997; Efstathiou \& Bond 1999).   If we take a
family of models with fixed initial perturbation spectra, fixed
physical densities $\wm \equiv \Omega_m h^2,  \wb \equiv \Omega_b
h^2$, 
and vary both $\Oml$ and the curvature $\Omega_k$ to keep a fixed value of 
the angular size distance to last scattering, then the
resulting CMB power spectra are identical (except for the integrated
Sachs-Wolfe effect at low multipoles which is hidden in cosmic
variance, and second-order effects at high $\ell$).
This degeneracy occurs because the physical densities $\wm, \wb$
control the structure of the perturbations in physical Mpc at last
scattering, while curvature 
and $\Lambda$ (plus $\wm$) govern the proportionality
between length at last scattering and observed angle.
Note that $h$ is a `derived' parameter in the above set, via $h = [
\wm / (1 - \Omega_k -\Oml) ]^{0.5}$, so the geometrical degeneracy is
broken by an external measurement of $h$.

\subsection{The flat-universe case} 

Assuming a flat universe, $\Omega_k = 0$, thus also
breaks the geometrical degeneracy.  However, as noted
by e.g. Efstathiou \& Bond (1999), and investigated below, there is a
closely related degeneracy based on varying two free parameters
(chosen from $\Omega_m, \wm, h, \Oml$) so as to almost preserve the
locations of the first few CMB acoustic peaks.  This is illustrated in
Fig~\ref{fig:ommvsh}, where the likelihood contours in the $(\Omega_m, h)$
plane for CMB-only data form a long and narrow `banana' with its long
axis at approximately constant $\Omega_m h^{3}$.  The banana is
surprisingly narrow in the other direction; this means that $\Omega_m
h^{3}$ is determined to about $12$\% (1$\sigma$) by the CMB data.
 
This `banana' is similar in form to the line in Fig.~4 of Efstathiou
\& Bond (1999), though different in detail because they used
simulations with $\wm = 0.25$.  It is also similar to that in Fig.~1
of Knox et~al. (2001) as expected.  However both those previous papers
presented the degeneracy in the $(\Oml, h)$ plane; although this is
just a mirror-image of the $(\Omega_m, h)$ plane, it is less intuitive
(e.g. changing $\Oml$ alters observables  that have no explicit
$\Lambda$-dependence, via the constraint $\Omega_m = 1 - \Oml$), so
the simple $\Omega_m h^3$ dependence has not been widely recognised.

\subsection{Peak locations and the sound horizon} 

The locations $\ell_m$ of the first few CMB acoustic peaks may be
 conveniently expressed (e.g. Hu et~al. 2001, Knox et~al. 2001) as
\begin{equation} 
\label{eq:la} 
\ell_m = \ell_{\japsub A} (m-\phi_m), \qquad m = 1,2,3
\end{equation}  
\begin{equation} 
  \ell_{\japsub A} \equiv \pi / \theta_{\japsub S} 
\end{equation}  
\begin{equation} 
\label{eq:thetas} 
  \theta_{\japsub S} \equiv {r_{\japsub S}(z_*) \over D_{\japsub A}(z_*)},   
\end{equation}  
where $r_{\japsub S}$ is the sound horizon size at last scattering 
(redshift $z_*$), $D_{\japsub A}$ is 
the angular diameter distance to last scattering,
therefore $\theta_{\japsub S}$ is the `sound horizon angle' 
and $\ell_{\japsub A}$ is the
`acoustic scale'.  For any given model, the CMB peak locations $\ell_m
(m=1,2,3)$ are given by numerical computation, and then
Eq.~(\ref{eq:la}) defines the empirical `phase shift' parameters
$\phi_m$. Hu et~al. (2001) show that the $\phi_m$'s are weakly
dependent on cosmological parameters and $\phi_1 \sim 0.27$, $\phi_2
\sim 0.24$, $\phi_3 \sim 0.35$.  Extensive calculations of the
$\phi_m$'s are given by Doran \& Lilley (2002).

Therefore, although $\theta_{\japsub S}$ is not directly observable, it is very
simple to compute and very tightly related to the peak locations,
hence its use below.  Knox et~al. (2001) note a `coincidence' that
$\theta_{\japsub S}$ is tightly correlated 
with the age of the universe for flat
models with parameters near the `concordance' values, and use this to
obtain an accurate age estimate assuming flatness.
 
\subsection{A heuristic explanation} 

Here we provide a simple heuristic explanation for why $\theta_{\japsub S}$ and
hence the $\ell_m$'s are primarily dependent on the parameter
combination $\Omega_m h^{3.4}$.

Of the four `FRW' parameters $\Omega_m, \wm, h, \Oml$, only 2 are
independent for flat models, and we can clearly choose any pair except
for $(\Omega_m, \Oml)$.  The standard choice in CMB analyses is $(\wm,
\Oml)$ while for non-CMB work the usual choice is $(\Omega_m, h)$.
However in the following we take $\wm$ and $\Omega_m$ to be the
independent parameters (thus $h, \Oml$ are derived); this looks
unnatural but separates more clearly the low-redshift effect
 of $\Omega_m$ from the pre-recombination effect of $\wm$.  We take
$\wb$ as fixed unless otherwise specified (its effect here is small).
 
We first note that the present-day horizon size for flat models is
well approximated by (Vittorio \& Silk 1985)
\begin{equation}
 \label{eq:rhnow} 
  r_{\japsub H}(z=0) = \frac{2c}{H_0}\Omega_m^{-0.4} = 6000 {\,\rm Mpc} 
 \, {\Omega_m^{0.1}  \over \sqrt{\wm} . } 
\end{equation}
(The distance to last scattering is $\sim 2\%$ smaller than the
above due to the finite redshift of last scattering).  Therefore, if
we increase $\Omega_m$ while keeping $\wm$ fixed, the shape and
relative heights of the CMB peaks are preserved but the peaks move
slowly rightwards (increasing $\ell$) proportional to
$\Omega_m^{0.1}$  (Equivalently, the Efstathiou-Bond $\cal R$
parameter for flat models is well approximated by $1.94\,
\Omega_m^{0.1}$).

This slow variation of 
$\ell_{\japsub A} \propto \Omega_m^{0.1}$ at fixed $\wm$ explains why 
the geometrical degeneracy is only weakly broken by the restriction to
flat models: a substantial change in $\Omega_m$ at fixed $\wm$ moves
the peaks only slightly, so a small change in $\wm$ can alter the
sound horizon length $r_{\japsub S}(z_*)$ and bring the peaks back to their
previous angular locations with only a small change in relative
heights.  We now give a simplified argument for the dependence of
$r_{\japsub S}$ on $\wm$.

The comoving sound horizon size at last scattering is defined by
(e.g. Hu \& Sugiyama 1995)
\begin{equation}
\label{eq:rs} 
 r_{\japsub S}(z_*) \equiv \frac{1}{H_0 \Omega_m^{1/2}} 
   \int_0^{a_*} {c_{\japsub S} \over (a + a_{\rm eq})^{1/2} } \, da 
\end{equation} 
where vacuum energy is neglected at these high redshifts;
the expansion factor $a \equiv (1+z)^{-1}$ and
$a_*, \aeq$ are the values at recombination and
matter-radiation equality respectively.
 Thus $r_{\japsub S}$ depends on $\wm$ and
$\wb$ in several ways:
\begin{description}  
\item[(i)] The expansion rate in the denominator depends on $\wm$ 
 via $\aeq$.  
\item[(ii)] The sound speed $c_{\japsub S}$ depends on the baryon/photon ratio
via $c_{\japsub S} = c / \sqrt{3(1+R)}$, $R = 30496 \,\wb \, a$.
\item [(iii)] The recombination redshift $z_*$ depends on both the
 baryon and matter densities $\wb, \wm$ in a fairly complex way.
\end{description} 

Since we are interested mainly in the {\em derivatives} of $r_{\japsub S}$ with
$\wm,~\wb$, it turns out that (i) above is the dominant effect.  The
dependence (iii) of $z_*$ on $\wm, \wb$ is slow. 
Concerning (ii), for baryon densities
$\wb \simeq 0.02$, $c_{\japsub S}$ declines smoothly from $c/\sqrt{3}$ at high
redshift to $0.80 \, c/\sqrt{3}$ at recombination.  Therefore to a
reasonable approximation we may take a fixed `average' $c_{\japsub S} \simeq
0.90 \, c / \sqrt{3}$ outside the integral in Eq.~(\ref{eq:rs}), and take
a fixed $z_*$, giving the approximation
\begin{equation}  \label{eq:rsapprox} 
  r_{\japsub S}(z_*) \simeq {0.90 \over \sqrt{3}} \, r_{\japsub H} (z = 1100) 
\end{equation} 
where $r_{\japsub H}$ is the light horizon size; this approximation 
 is very accurate for all
$\wm$ considered here and $\wb \simeq 0.02$.  For other baryon
densities, multiplying the rhs of Eq.~(\ref{eq:rsapprox}) by $(\wb/0.02)^{-0.07}$
is a refinement.  (Around the concordance value $\wb = 0.02$, effects
(ii) and (iii) partly cancel, because increasing $\wb$ lowers the
sound speed but also delays recombination i.e. increases $a_*$). 
 
From above, the (light) horizon size at recombination is 
\begin{equation}
 \label{eq:rhrec} 
  r_{\japsub H}(z_*) = \frac{c}{H_0 \Omega_m^{1/2}} \int_0^{a_*} {1 \over (a +
  \aeq )^{1/2} } \, da 
\end{equation}
\begin{displaymath}
 \qquad   = {6000 \,{\rm Mpc} \over \sqrt{\wm}} 
  \sqrt{a_*} \left[\sqrt{1 + (\aeq /a_*)} 
  - \sqrt{\aeq / a_*} \right] 
\end{displaymath}
Dividing by $D_{\japsub A} \simeq 0.98\, r_{\japsub H}(z=0)$ from Eq.~(\ref{eq:rhnow}) 
 gives the angle subtended today by
the light horizon, 
\begin{equation}
  \theta_{\japsub H} \simeq 1.02\, \frac{\Omega_m^{-0.1}}{\sqrt{1+z_*}}
     \left[\sqrt{1 + \frac{\aeq}{a_*} } - 
  \sqrt{\frac{\aeq}{a_*} }\, \right].
\end{equation}
Inserting $z_* = 1100$ and $a_{\rm eq} = (23900 \,\wm)^{-1}$,
we have
\begin{equation} 
 \label{eq:thetah} 
\eqalign{
 \theta_{\japsub H} &= \frac{1.02 \, \Omega_m^{-0.1}}{\sqrt{1101}} \quad \times \cr
 &\left[\sqrt{1 + 0.313 \left({0.147 \over \wm}\right)} 
 - \sqrt{0.313 \left({0.147 \over \wm}\right) } \right] , \cr
} 
\end{equation} 
and $\theta_{\japsub S} \simeq \theta_{\japsub H} \times 0.9 / \sqrt{3} $ from
Eq.~(\ref{eq:rsapprox}). 
 This remarkably simple result captures  most
of the parameter dependence of CMB peak locations within
flat $\Lambda$CDM  models. 
Note that the square bracket in Eq.~(\ref{eq:thetah}) tends (slowly) 
 to 1 for  $a_{\rm eq} \ll a_*$ i.e. $\wm \gg 0.046$; thus 
 it is the fact that matter
 domination was not much earlier than recombination which leads to
 the significant dependence of 
 $\theta_{\japsub H}$ on $\wm$ and hence $h$. 

Differentiating Eq.~(\ref{eq:thetah}) near a fiducial $\wm = 0.147$ gives  
\begin{displaymath}
  \left.\frac{\partial\ln\theta_{\japsub
  H}}{\partial\ln\Omega_m}\right|_\wm 
   = -0.1,
\end{displaymath} 
\begin{equation} 
  \label{eq:deriv} 
  \left.\frac{\partial\ln\theta_{\japsub
  H}}{\partial\ln\wm}\right|_{\Omega_m}=
  \frac{1}{2} \left( 1 + \frac{a_*}{\aeq}\right)^{-1/2} =  +0.24 ,
\end{equation}  
and the same for derivatives of $\ln \theta_{\japsub S}$ from the approximation
above.
In terms of $(\Omega_m, h)$ this gives
\begin{equation}
  \left.\frac{\partial\ln\theta_{\japsub H}}
 {\partial\ln\Omega_m}\right|_h =+0.14,
  \hspace{1cm}
  \left.\frac{\partial\ln\theta_{\japsub H}}{\partial\ln h}\right|_{\Omega_m}=+0.48 ,
\end{equation}  
in good agreement with the numerical derivatives of $\ell_{\japsub A}$ in
Eq.~(A15) of Hu et~al. (2001).  Note also the sign difference between
the two $\partial / \partial \ln \Omega_m$ values above.

Thus for moderate variations from a `fiducial' model, the CMB peak
locations scale approximately as $\ell_m \propto \Omega_m^{-0.14}
h^{-0.48}$, i.e.  the condition for constant CMB peak location is well
approximated as $\Omega_m h^{3.4} = {\rm constant}$. This condition
can also be written $\wm\Omega_m^{-0.41} = {\rm constant}$, and we see
that, along such a contour, $\wm$ varies as $\Omega_m^{0.41}$, and
hence the peak heights are slowly varying and the overall CMB power
spectrum is also slowly varying.

\begin{figure}
  \setlength{\epsfxsize}{0.95\columnwidth} \centerline{\epsfbox{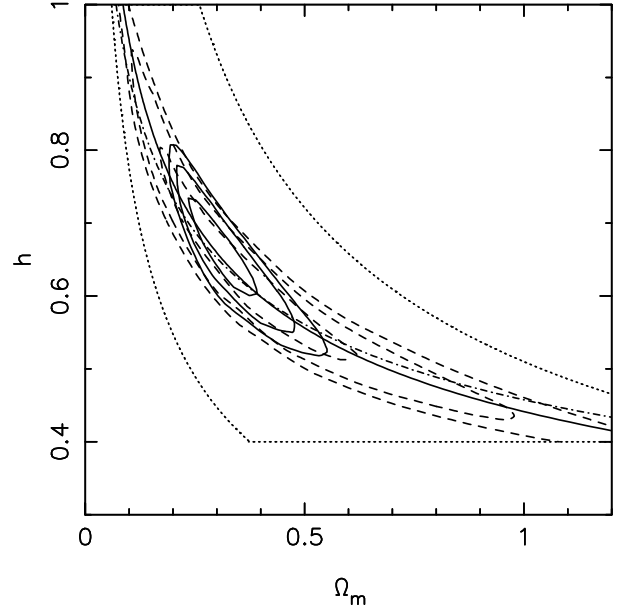}}

  \caption{Likelihood contours for $\Omega_m$ against $h$ for scalar
  only models, plotted as in Fig.~\ref{fig:like_sca}. Variables were
  changed from $\Omega_bh^2$ and $\Omega_ch^2$ to $\Omega_m$ and
  $\Omega_b/\Omega_m$, and a uniform prior was assumed for
  $\Omega_b/\Omega_m$ covering the same region as the original
  grid. The extent of the grid is shown by the dotted lines. The
  dot-dash line follows the locus of models through the likelihood
  maximum with constant $\Omega_mh^{3.4}$. The solid line is a fit to
  the likelihood valley and shows the locus of models with constant
  $\Omega_mh^{3.0}$ (see text for details).\label{fig:ommvsh}}
\end{figure}

\begin{figure*}
  \setlength{\epsfxsize}{0.65\textwidth} 
  \centerline{\epsfbox{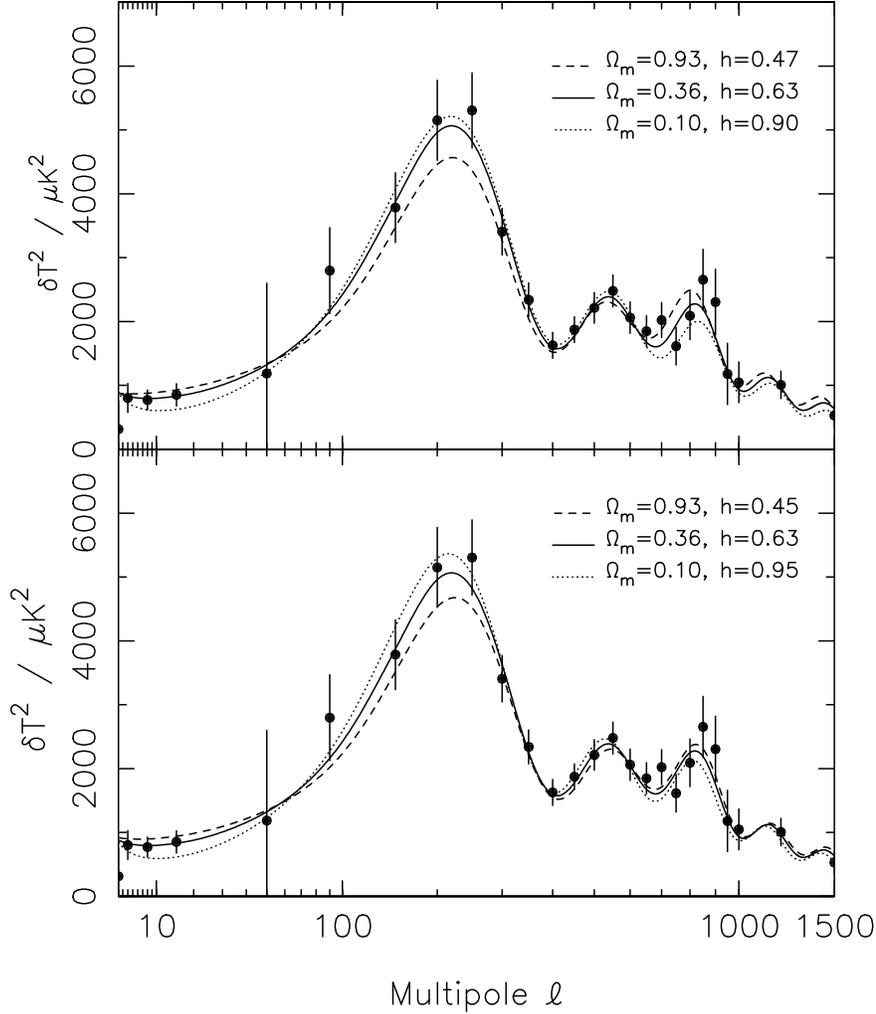}} 

 \caption{The top panel shows three scalar only model CMB angular
  power spectra with the same apparent horizon angle, compared to the
  data of Table~\ref{tab:calib}.  Although these models have
  approximately the same value of $\Omega_m h^{3.4}$, they are
  distinguishable by peak heights. Such additional constraints alter
  the degeneracy observed in Fig.~\ref{fig:ommvsh} slightly from
  $\Omega_m h^{3.4}$ to $\Omega_m h^{3.0}$. Three scalar only models
  that lie in the likelihood ridge with $\Omega_m h^{3.0}$ are
  compared with the data in the bottom panel. For all of the models
  shown, parameters other than $\Omega_m$ and $h$ have been adjusted
  to their maximum-likelihood positions.  \label{fig:cls_valley}}
\end{figure*}

There are four approximations used for $\theta_{\japsub S}$ 
 above: one in Eq.~(\ref{eq:rhnow}),
two (constant $c_s$ and $z_*$) in Eq.~(\ref{eq:rsapprox}),
and finally the $\Omega_m h^{3.4}$ line is a first-order (in log) 
 approximation to a contour of constant Eq.~(\ref{eq:thetah}). 
Checking against numerical results, we find that  each of these 
causes up to 1\% error in $\theta_{\japsub S}$, but
 they partly cancel:  the exact value of $\theta_{\japsub S}$
 varies by $< 0.5\%$ along the contour $h = 0.7 \,
 (\Omega_m/0.3)^{-1/3.4}$ between $0.1 \le \Omega_m \le 1$. 
The peak heights shift the numerical 
degeneracy by more than this (see below), so the error is unimportant.  

A line of constant $\Omega_m h^{3.4}$ is compared to the likelihood
surface recovered from the CMB data in Fig.~\ref{fig:ommvsh}. In order
to calculate the required likelihoods, we have made a change of
variables from $\wb$ \& $\Omega_c h^2$ to $\Omega_m$ \&
$\Omega_b/\Omega_m$, and have marginalized over the baryon fraction
assuming a uniform prior in $\Omega_b/\Omega_m$ covering the limits of
the grid used. As expected, the degeneracy observed when fitting the
CMB data alone is close to a contour of constant $\ell_{\japsub A}$
hence constant $\theta_{\japsub H}$. However, information about the
peak heights does alter this degeneracy slightly; the relative peak
heights are preserved at constant $\wm$, hence the actual likelihood
ridge is a `compromise' between constant peak location (constant
$\Omega_m h^{3.4}$) and constant relative heights (constant $\Omega_m
h^2$); the peak locations have more weight in this compromise, leading
to a likelihood ridge along $\Omega_m h^{3.0} \simeq {\rm const}$.
This is shown by the solid line in Fig.~\ref{fig:ommvsh}. To
demonstrate where this alteration is coming from, we have plotted
three scalar only models in the top panel of
Fig.~\ref{fig:cls_valley}. These models lie approximately along the
line of constant $\Omega_m h^{3.4}$, with $\Omega_m=0.93,0.36,0.10$.
Parameters other than $\Omega_m$ and $h$ have been adjusted to fit to
the CMB data. The differing peak heights (especially the 3rd peak)
between the models are clear (though not large) and the data therefore
offer an additional constraint that slightly alters the observed
degeneracy. The bottom panel of Fig.~\ref{fig:cls_valley} shows three
models that lie along the observed degeneracy, again with
$\Omega_m=0.93,0.36,0.11$. The narrow angle of intersection between
contours of constant $\Omega_m h^{3.4}$ and $\Omega_m h^2$ (only 10
degrees in the $(\ln \Omega_m, \ln h)$ plane) explains why the
likelihood banana is long.

The exponent of $h$ for constant $\theta_{\japsub H}$ varies slowly
from 2.9 to 4.1 as $\wm$ varies from 0.06 to 0.26. 
Note that Hu et~al. (2001) quote an exponent of 3.8 for constant
$\ell_1$; the difference from 3.4 is mainly due to the slight
dependence of $\phi_1$ on $\wm$ which we ignored above.  However since
that paper, improved CMB data has better revealed the 2nd and 3rd
peaks, and the exponent of $3.4$ is more appropriate for preserving
the mean location of the first 3 peaks. Also, note that the
near-integer exponent of 3.0 for the likelihood ridge in
Fig.~\ref{fig:ommvsh} is a coincidence that depends on the 
 observed value of $\wm$, details of the CMB error bars
etc. However, the arguments above are fairly generic, so we anticipate
that any CMB dataset covering the first few peaks should 
(assuming flatness) 
give a likelihood ridge elongated along a contour of  
 constant $\Omega_m h^p$, with $p$ fairly close to 3.  

To summarise this section, the CMB peak locations are closely related
to the angle subtended by the sound horizon at recombination, which we
showed is a near-constant fraction of the light horizon angle given in
Eq.~(\ref{eq:thetah}).  We have thus called this the `horizon angle
degeneracy' which has more physical content than the alternative `peak
location degeneracy'.  A contour of constant $\theta_{\japsub S}$ is
very well approximated by a line of constant $\Omega_m h^{3.4}$, and
information on the peak heights slightly `rotates' the measured
likelihood ridge near to a contour of constant $\Omega_m h^{3.0}$.

\section{Constraining quintessence}  \label{sec:quin}

\begin{figure}
  \setlength{\epsfxsize}{0.95\columnwidth} \centerline{\epsfbox{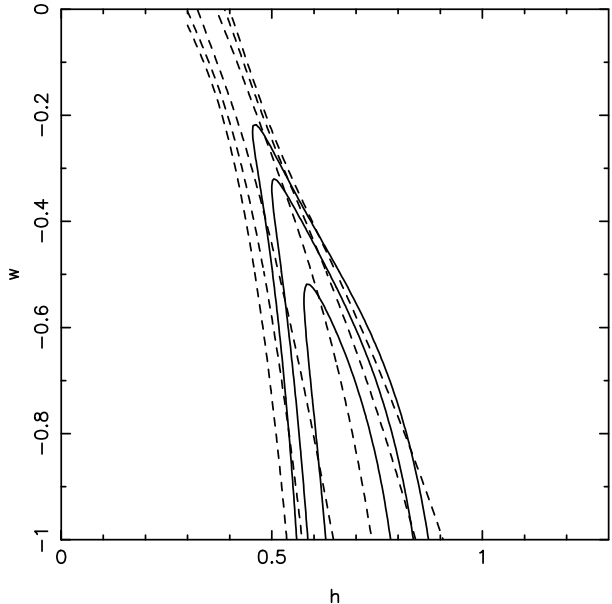}}

  \caption{Likelihood contours for the equation of state of the vacuum
  energy parameter $w$ against the Hubble constant $h$. Dashed
  contours are for CMB+2dFGRS data, solid contours include also the
  HST key project constraint.  Contours correspond to changes in the
  likelihood from the maximum of $2\Delta\ln{\cal L}=2.3, 6.0, 9.2$.
  Formally, this results in a 95\% confidence limit of $w<-0.52$.}
  \label{fig:w}
\end{figure}

There has been recent interest in a possible extension of the standard
cosmological model that allows the equation of state of the vacuum
energy $w \equiv p_{\rm vac} /\rho_{\rm vac}c^2$ to have $w\neq-1$
(e.g. Zlatev, Wang \& Steinhardt 1999), thereby not being a
`cosmological constant', but a dynamically evolving component. In this
section we extend our analysis to constrain $w$; we assume $w$ does
not vary with time since a model with time-varying $w$ generally looks
very similar to a model with suitably-chosen constant $w$ (e.g. Kujat
et~al. 2002).  The shapes of the CMB and matter power spectra are
invariant to changes of $w$ (assuming vacuum energy was negligible
before recombination): the only significant effect is to alter the
angular diameter distance to last scattering, and move the angles at
which the acoustic peaks are seen. For flat models, a useful
approximation to the present day horizon size is given by
\begin{equation}
 \label{eq:rhw} 
  r_{\japsub H}(z=0) = \frac{2c}{H_0}\Omega_m^{-\alpha}, 
  \qquad \alpha=\frac{-2w}{1-3.8w}
\end{equation}
(compare with Eq.~\ref{eq:rhnow} for $w=-1$). As discussed previously,
the primary constraint from the CMB data is on the angle subtended
today by the light horizon (given for $w=-1$ models by
Eq.~\ref{eq:thetah}). If $w$ is increased from $-1$ at fixed
$\Omega_m,h$, the peaks in the CMB angular power spectrum move to
larger angles.  To continue to fit the CMB data, we must decrease
$\Omega_m h^{3.4}$ to bring $\theta_{\japsub H}$ back to its `best-fit' value.
However, the 2dFGRS power spectrum constraint limits $\Omega_m h =
0.20 \pm 0.03$, so to continue to fit both CMB+2dFGRS we must reduce
$h$.  

The CMB and 2dFGRS datasets alone therefore constrain a combination of
$w$ and $h$, but not both separately.  The dashed lines in
Fig.~\ref{fig:w} show likelihood contours for $w$ against $h$ fitting
to both the CMB and 2dFGRS power spectra showing this effect. Here, we
have marginalized over $\Omega_m$ assuming a uniform prior
$0.0<\Omega_m<1.3$. An extra constraint on $h$ can be converted into a
limit on $w$: if we include the measurement $h = 0.72\pm 0.08$ from
the HST key project (Freedman et~al. 2001) we obtain the solid
likelihood contours in Fig.~\ref{fig:w}. The combination of these
three data sets then gives $w<-0.52$ (95\% confidence); the limit of
the range considered, $w=-1.0$, provides the smallest uncertainty.
The 95\% confidence limit is comparable to the $w<-0.55$ obtained from
the supernova Hubble diagram plus flatness (Garnavich et al. 1998).
See also Efstathiou (1999), who obtained $w<-0.6$ from a
semi-independent analysis combining CMB and supernovae (again assuming
flatness).

\section{Predicting the CMB power spectrum}  \label{sec:cls}

\begin{figure*}
  \setlength{\epsfxsize}{0.65\textwidth} \centerline{\epsfbox{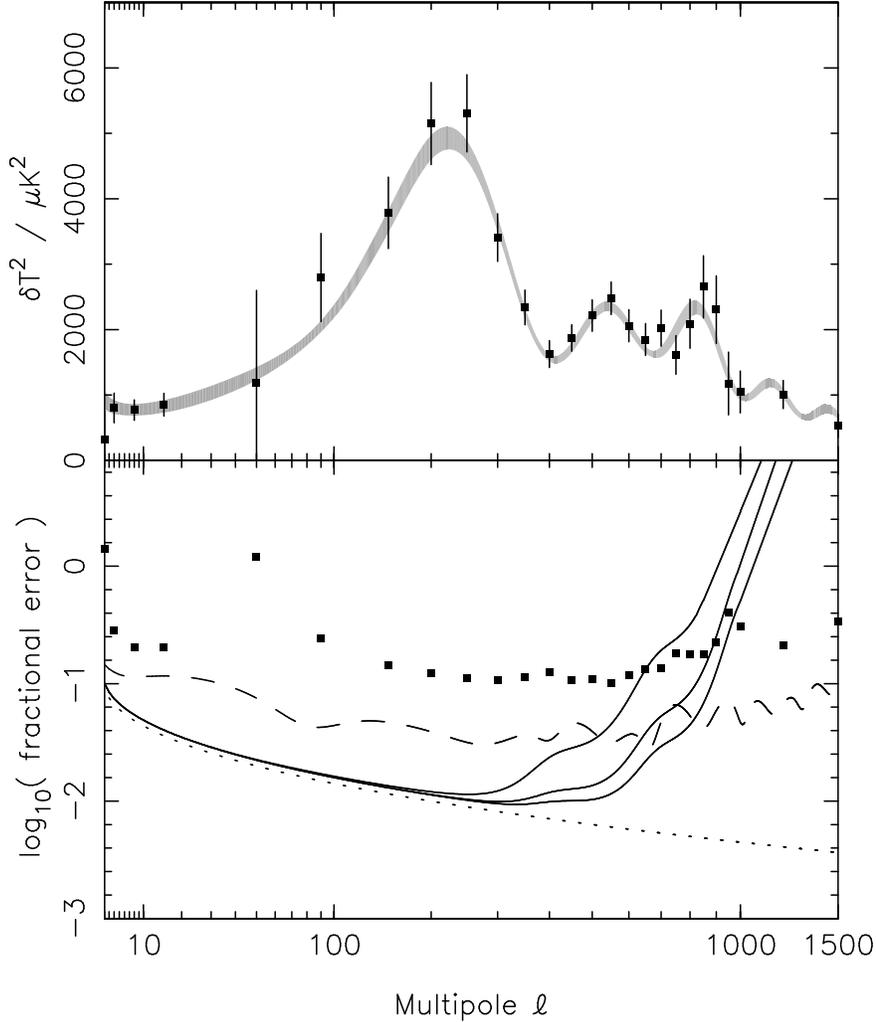}}

  \caption{Upper panel: the grey shaded region shows our prediction of
  the CMB angular power spectrum with 1$\sigma$ errors (see text);
  points show the data of Table~\ref{tab:calib}.  The lower panel
  shows fractional errors: points are the current data, dashed line is
  the errors on our prediction, and the three solid lines are expected
  errors for the MAP experiment (Page 2002) for $\Delta \ell=50$ and
  the 6 month, 2 year and 4 year data (top--bottom).  The dotted line
  shows the expected cosmic variance, again for $\Delta \ell=50$,
  assuming full sky coverage (the MAP errors assume 80\% coverage).
  As can be seen, the present CMB and LSS data provide a strong
  prediction over the full $\ell$-range covered by MAP.
  \label{fig:cls} }
\end{figure*}

An interesting aspect of this analysis is that the current CMB data
are rather inaccurate for $20 \simlt \ell \simlt 100$, and yet the
allowed CDM models are strongly constrained. We therefore consider how
well this model-dependent determination of the CMB power spectrum is
defined, in order to see how easily future data could test the basic
CDM+flatness paradigm.
 
\begin{table} 

  \caption{The predicted mean and rms CMB power calculated by
integrating the CMB+2dFGRS likelihood over the range of parameters
presented in Table~\ref{tab:grid}, allowing for a possible tensor
component. These data form a testable prediction of the CDM+flatness
paradigm. \label{tab:cls_predict} }

  \centering \begin{tabular}{ccc} \hline
  $\ell$ & $\delta T^2$ / $\mu K^2$ & rms error / $\mu K^2$ \\
  \hline
        2   &     920  & 134  \\
        4   &     817  & 102  \\
        8   &     775  &  88  \\
        15  &     828  &  96  \\
        50  &    1327  & 100  \\
        90  &    2051  &  87  \\
        150 &    3657  & 172  \\
        200 &    4785  & 186  \\
        250 &    4735  & 150  \\
        300 &    3608  & 113  \\
        350 &    2255  &  82  \\
        400 &    1567  &  52  \\
        450 &    1728  &  79  \\
        500 &    2198  &  86  \\
        550 &    2348  &  74  \\
        600 &    2052  &  76  \\
        650 &    1693  &  51  \\
        700 &    1663  &  79  \\
        750 &    1987  & 131  \\
        800 &    2305  & 122  \\
        850 &    2257  &  93  \\
        900 &    1816  & 107  \\
        950 &    1282  &  93  \\
        1000 &    982  &  46  \\
        1200 &   1029  &  69  \\
        1500 &    686  &  60  \\
  \hline
  \end{tabular}

\end{table}

Using our grid of $\sim2\times10^8$ models, we have integrated the
CMB+2dFGRS likelihood over the range of parameters presented in
Table~\ref{tab:grid} in order to determine the mean and rms CMB power
at each $\ell$. These data are presented in
Table~\ref{tab:cls_predict} at selected $\ell$ values, and the range
of spectra is shown by the grey shaded region in the top panel of
Fig.~\ref{fig:cls}. A possible tensor component was included in this
analysis, although this has a relatively minor effect, increasing the
errors slightly (as expected when new parameters are introduced), but
hardly affecting the mean values. The predictions are remarkably
tight: this is partly because combining the peak-location constraint
on $\Omega_m h^{3.4}$ with the 2dFGRS constraint on $\Omega_m h$ gives
a better constraint on $\Omega_m h^2$ than the CMB data alone, and
this helps to constrain the predicted peak heights.

The bottom panel of Fig.~\ref{fig:cls} shows the errors on an
expanded scale, compared with the cosmic variance limit and the
predicted errors for the MAP experiment (Page 2002).  This comparison
shows that, while MAP will beat our present knowledge of the CMB
angular power spectrum for all $\ell\simlt 800$, this will be
particularly apparent around the first peak.  As an example of the
issues at stake, the scalar-only models predict that the location of
the first CMB peak should be at $\ell=221.8 \pm 2.4$.  Significant
deviations observed by MAP from such predictions will imply that some
aspect of this model (or the data used to constrain it) is
wrong. Conversely, if the observations of MAP are consistent with this
band, then this will be strong evidence in favour of the model.

\section{Summary and discussion}

Following recent releases of CMB angular power spectrum measurements
from VSA and CBI, we have produced a new compilation of data that
estimates the true power spectrum at a number of nodes, assuming that
the power spectrum behaves smoothly between the nodes. The best-fit
values are not convolved with a window function, although they are not
independent. The data and Hessian matrix are available from {\tt
http://www.roe.ac.uk/{\tt\char'176}wjp/CMB/}. We have used these data
to constrain a uniform grid of $\sim2\times10^8$ flat cosmological
models in 7 parameters jointly with 2dFGRS large scale structure
data. By fully marginalizing over the remaining parameters we have
obtained constraints on each, for the cases of CMB data alone, and
CMB+2dFGRS data. The primary results of this paper are the resulting
parameter constraints, particularly the tight constraints on $h$ and
the matter density $\Omega_m$: combining the 2dFGRS power spectrum
data of Percival et~al. (2001) with the CMB data compilation of
Section~\ref{sec:CMBdata}, we find $h=0.665\pm0.047$ and
$\Omega_m=0.313\pm0.055$ (standard rms errors), for scalar-only
models, or $h=0.700\pm0.053$ and $\Omega_m=0.275\pm0.050$, allowing a
possible tensor component.

We have also discussed in detail how these parameter constraints
arise.  Constraining $\Omega_{\rm tot}=1$ does not fully break the
geometrical degeneracy present when considering models with varying
$\Omega_{\rm tot}$, and models with CMB power spectra that peak at the
same angular position remain difficult to distinguish using CMB data
alone. A simple derivation of this degeneracy was presented, and
models with constant peak locations were shown to closely follow lines
of constant $\Omega_m h^{3.4}$. We can note a number of interesting
phenomenological points from this analysis:

\begin{enumerate} 
\item The narrow CMB $\Omega_m-h$ likelihood ridge in
Fig.~\ref{fig:ommvsh} derives primarily from the peak {\em locations},
therefore it is insensitive to many of the parameters affecting peak
{\em heights}, e.g. tensors, $n_s$, $\tau$, calibration uncertainties
etc. Of course it is strongly dependent on the flatness assumption.
\item This simple picture is broken in detail as the current CMB data
obviously place additional constraints on the peak heights. This
changes the degeneracy slightly, leading to a likelihood ridge near
constant $\Omega_m h^{3}$.
\item The high power of $h^{3}$ means that adding an external $h$
constraint is not very powerful in constraining $\Omega_m$, but an
external $\Omega_m$ constraint gives strong constraints on $h$.  A
10\% measurement of $\Omega_m$ (which may be achievable e.g. from 
evolution of cluster abundances) would give a 4\% measurement of $h$.
\item When combined with the 2dF power spectrum shape (which mainly
constrains $\Omega_m h$), the CMB+2dFGRS data gives a constraint on
 $\Omega_m h^2 = 0.1322\pm0.0093$ (including tensors)
or $\Omega_m h^2 = 0.1361\pm0.0096$ (scalars only), which is
considerably tighter from the CMB alone. Subtracting the baryons
gives $\Omega_c h^2 = 0.1096\pm0.0092$ (including tensors) or
$\Omega_c h^2 = 0.1151\pm0.0091$ (scalars only), accurate results
that may be valuable in constraining the parameter space of particle
dark matter models and thus predicting rates for 
 direct-detection experiments. 
\item We can understand the solid contours in Figure~\ref{fig:ommvsh}
simply as follows: the CMB constraint can be approximated as a
1-dimensional stripe $\Omega_m h^{3.0} = 0.0904\pm0.0092$ (including
tensors) or $\Omega_m h^{3.0} = 0.0876\pm0.0085$ (scalars only),
and the 2dF constraint as another stripe $\Omega_m h = 0.20 \pm
0.03$. Multiplying two Gaussians with the above parameters gives a
result that looks quite similar to the fully-marginalized contours.
In fact, modelling the CMB constraint simply using the location of the
peaks to give $\Omega_m h^{3.4} = 0.081\pm0.012$ (including tensors)
or $\Omega_m h^{3.4} = 0.073\pm0.010$ (scalars only) also
produces a similar result, demonstrating that the primary constraint
of the CMB data in the $(\Omega_m, h)$ plane is on the apparent
horizon angle.
\end{enumerate}  

In principle, accurate non-CMB measurements of both $\Omega_m$ and $h$
can give a robust prediction of the peak locations assuming
flatness. If the observed peak locations are significantly different,
this would give evidence for either non-zero curvature, quintessence
with $w \neq -1$ or some more exotic failure of the model.  Using the
CMB data to constrain the horizon angle, and 2dFGRS data to constrain
$\Omega_mh$, there remains a degeneracy between $w$ and $h$. This can
be broken by an additional constraint on $h$; using $h=0.72\pm0.08$
from the HST key project (Freedman et~al. 2001), we find $w<-0.52$ at
95\% confidence. This result is comparable to that found by Efstathiou
(1999) who combined the supernovae sample of Perlmutter et~al. (1999)
with CMB data to find $w<-0.6$.

In Section~\ref{sec:cls} we considered the constraints that combining
the CMB and 2dFGRS data place on the CMB angular power spectrum. This
was compared with the predicted errors from the MAP satellite in order
to determine where MAP will improve on the present data and provide
the strongest constraints on the cosmological model. It will be
fascinating to see whether MAP rejects these predictions, thus
requiring a more complex cosmological model than the simplest flat
CDM-dominated universe.

Finally, we announce the public release of the 2dFGRS power spectrum
data and associated covariance matrix determined by Percival et~al.
(2001). We also
provide code for the numerical calculation of the convolved power
spectrum and a window matrix for the fast calculation of the convolved
power spectrum at the data values.
The data are available from either {\tt
http://www.roe.ac.uk/{\tt\char'176}wjp/} or from {\tt
http://www.mso.anu.edu.au/2dFGRS}; as we have demonstrated, they are a
critical resource for constraining cosmological models. 

\section*{ACKNOWLEDGEMENTS}

The 2dF Galaxy Redshift Survey was made possible through the dedicated
efforts of the staff of the Anglo-Australian Observatory, both in
creating the 2dF instrument and in supporting it on the telescope.

\end{document}